\documentstyle[12pt,epsf]{article}

\def\beq{\begin{equation}}
\def\eeq{\end{equation}}
\def\beqar{\begin{eqnarray}}
\def\eeqar{\end{eqnarray}}

\def\he#1{\hbox{${}^{#1}$He}}
\def\li#1{\hbox{${}^{#1}$Li}}

\def\yp{\hbox{$Y_{\rm p}$}}

\def\la{\mathrel{\mathpalette\fun <}}
\def\ga{\mathrel{\mathpalette\fun >}}
\def\fun#1#2{\lower3.6pt\vbox{\baselineskip0pt\lineskip.9pt
  \ialign{$\mathsurround=0pt#1\hfil##\hfil$\crcr#2\crcr\sim\crcr}}}
\begin{document}

\def\lsim{\mathrel{\vcenter{\hbox{$<$}\nointerlineskip\hbox{$\sim$}}}}
\def\gsim{\mathrel{\vcenter{\hbox{$>$}\nointerlineskip\hbox{$\sim$}}}}
\def\erf{\mathop{\rm erf}\nolimits}
\def\nnu{N_\nu}% Use in math mode
\def\Hunit{km s$^{-1}$ Mpc$^{-1}$}
\def\ie{i.e.}
\def\eg{e.g.}
\def\etal{et al.}
\def\PLB{Phys.\ Lett.\ B}
\def\PRC{Phys.\ Rev.\ C}
\def\PRD{Phys.\ Rev.\ D}
\def\PRL{Phys.\ Rev.\ Lett.}
\def\MNRAS{Mon. Not. R. Astr. Soc.\ }
\def\ApJ{Ap.\ J.}
\def\ApJS{Ap.\ J.\ Suppl.\ Ser.}

\begin{titlepage}
\pagestyle{empty}
\baselineskip=15pt
\rightline{UMN-TH-1728/98}
\rightline{TPI-MINN-98/23}
\rightline{hep-ph/9811444}
\rightline{November 1998}
\vskip .2in
\baselineskip=15pt
\bigskip

\begin{center}
{\bf GENERALIZED LIMITS TO 
THE NUMBER OF LIGHT PARTICLE DEGREES OF FREEDOM 
FROM BIG BANG NUCLEOSYNTHESIS}
\end{center}
\renewcommand{\thefootnote}{\alph{footnote}}
\bigskip
\centerline{Keith A. Olive\footnote{olive@mnhep.hep.umn.edu}
}
\smallskip
\centerline{\it University of Minnesota, School of Physics and Astronomy}
\centerline{\it Theoretical Physics Institute}
\centerline{\it 116 Church St SE, Minneapolis, MN 55455}

\bigskip
\centerline{David Thomas\footnote{davet@oddjob.uchicago.edu}
}
\smallskip
\centerline{\it University of Chicago, Astronomy \& Astrophysics Center}
\centerline{\it 5640 S Ellis Ave, Chicago, IL 60637}
\bigskip

\centerline{\bf Abstract}

{\narrower
\noindent
We compute the big bang nucleosynthesis limit on the number of light 
neutrino degrees of freedom in a model-independent likelihood analysis 
based on the abundances of \he4 and \li7. 
We use the two-dimensional likelihood functions to simultaneously
constrain the baryon-to-photon ratio and
the number of light neutrinos for a range of \he4 abundances $Y_p$ = 0.225 --
0.250, as well as a range in primordial 
\li7 abundances from (1.6 to 4.1) $ \times 10^{-10}$. 
For (\li7/H)$_p = 1.6\times10^{-10}$, as can be inferred from the \li7
data from Population II halo stars, the upper limit to $\nnu$ based on the
current best estimate of the primordial \he4 abundance of $Y_p = 0.238$,
is $\nnu < 4.3$ and varies from
$\nnu<3.3$ (at 95\% C.L.) when $Y_p=0.225$ to $\nnu<5.3$ when
$Y_p=0.250$. If \li7 is depleted in these stars the upper limit to $\nnu$ is
relaxed.  Taking (\li7/H)$_p = 4.1\times10^{-10}$, the limit
varies from $\nnu<3.9$ when $Y_p = 0.225$ to $\nnu 
\la 6$ when $Y_p = 0.250$.
We also consider the consequences on the upper limit to $\nnu$
if recent observations of deuterium in high-redshift quasar
absorption-line systems are confirmed.
\par}
\end{titlepage}
\vfill\eject
\baselineskip=\normalbaselineskip
\setcounter{footnote}{0}
\section{Introduction}

One of the most important limits on particle properties is the limit on the
number of light particle degrees of freedom at the the time of big bang
nucleosynthesis (BBN) \cite{ssg}.  This is commonly computed as a limit on
the number of light neutrino flavors, $\nnu$. Recently, we \cite{oth2}
used a model-independent likelihood method (see also \cite{fo,fkot}) to
simultaneously constrain the value of the one true parameter in  {\em
standard} BBN, the baryon-to-photon ratio $\eta$, together with $\nnu$.
For similar approaches, see \cite{cst2}. In that work \cite{oth2}, we based
our results on the best estimate of the observationally determined
abundance of \he4, $Y_p = 0.234 \pm 0.002 \pm 0.005$ from \cite{ostsk},
and of \li7, \li7/H $= (1.6 \pm 0.1) \times 10^{-10}$, from \cite{mol}.
While these determinations can still be considered good ones today, there
is often discussion of higher abundance for \he4 as perhaps indicated by
the data of \cite{iz} and higher abundances of \li7 due to the effects of
stellar depletion (see e.g.
\cite{pinn}). Rather than be forced to continually update the limit on
$\nnu$ as the observational situation evolves, we generalize our previous
work here and compute the upper limit on $\nnu$ for a wide range of
possible observed abundances of \he4 and \li7.  Because the
determinations of D/H in quasar absorption system has not dramatically
improved, we can only comment on the implications of either the high or
low D/H measurements. 

One of the major obstacles in testing BBN using the observed abundances of the
light element isotopes rests on our ability to infer from these observations a
primordial abundance. Because \he4, in extragalactic
HII regions, and \li7, in the atmospheres of old halo dwarf stars, are both
measured in very low metallicity systems (down to 1/50th solar for \he4 and 1/1000th
solar for \li7), very little modeling in the way of galactic chemical
evolution is required to extract a primordial abundance for these
isotopes. Of course systematic uncertainties, such as underlying stellar
absorption, in determining the \he4 abundance and the effects of stellar
depletion of \li7 lead to uncertainties in the primordial abundances of these
isotopes, and it is for that reason we are re-examining the limits to $\nnu$.
Nevertheless, the problems in extracting a primordial \he4 and \li7 abundance
pale in comparison with those for D and \he3, both of which are subject to
considerable uncertainties not only tied to the observations, but to galactic
chemical evolution. In fact, \he3 also suffers from serious uncertainties
concerning its fate in low mass stars \cite{orstv}. \he3 is both
produced  and destroyed in stars making the connection to BBN very
difficult. 

Deuterium is totally destroyed in the star formation process.  As such, the
present or solar abundance of D/H is highly dependent on the details of a
chemical evolution model, and  in particular the galactic star formation rate.
Unfortunately, it is very difficult at the present time to gain insight on the
primordial abundance of D/H from chemical evolution given present and solar
abundances since reasonably successful models of chemical evolution can be
constructed for primordial D/H values which differ by nearly an order of
magnitude\footnote{There may be some indication from studies of the
luminosity density at high redshift which implies a steeply decreasing
star formation rate \cite{cce}, and that at least on a cosmic scale,
significant amounts of deuterium has been destroyed \cite{cova}.}
\cite{scov}.

Of course much of the recent excitement surrounding deuterium concerns the
observation of D/H in quasar absorption systems
\cite{quas1}-\cite{quas4}. If a single value for the D/H abundance in
these systems could be established\footnote{It is not possible that all
disparate determinations of D/H represent an inhomogeneous primordial
abundance as the corresponding inhomogeneity in $\eta$ would lead to
anisotropies in the microwave background in excess of those observed
\cite{cos2}.}, then one could avoid all of the complications concerning
D/H and chemical evolution, and because of the steep monotonic dependence
of D/H on $\eta$, a good measurement of D/H would alone be sufficient to
determine the value of $\eta$ (since D/H is nearly independent of $\nnu$).
In this case, the data from \he4 and \li7 would be most valuable as a
consistency test on BBN and in the case of \he4, to set limits on
particle properties. In the analysis that follows, we will discuss the
consequences of the validity of either the high or low D/H determinations.

Using a likelihood analysis based on \he4 and \li7 
\cite{fkot}, a probable range for the baryon-to-photon ratio, $\eta$ was
determined. 
 The \he4 likelihood distribution has a single
peak due to the monotonic dependence of \he4 on
$\eta$. However, because the dependence on $\eta$ is relatively flat,
particularly at higher values of $Y_p$, this peak may be very broad,
yielding little information on $\eta$ alone. On the other hand, because
\li7 is not monotonic in
$\eta$, the BBN prediction has  a minimum at $\eta_{10}
\simeq 3$ ($
\eta_{10} = 10^{10} \eta$) and as a result, for an observationally determined value
of \li7 above the minimum, the \li7 likelihood distribution will show two peaks. 
The total likelihood distribution
based on \he4 and \li7 is simply the product of the two individual
distributions.
In \cite{fkot}, the best fit value for
$\eta_{10}$ based on the quoted observational abundances was found to be 
1.8 with a 95\% CL range 
\beq
1.4  <  \eta_{10}  <   4.3  
\label{etar}
\eeq
when restricting the analysis to the
standard model, including $\nnu = 3$.
In determining (\ref{etar}) systematic errors were treated as Gaussian
distributed.
When D/H from quasar absorption systems (those showing a 
high value for D/H \cite{quas1,quas3}) is included in the analysis this
range is cut to $1.50 < \eta_{10} < 2.55$.

In \cite{oth2}, the maximum likelihood analysis of \cite{fo,fkot}  which
utilized a likelihood function ${\cal L}(\eta)$ for fixed $\nnu = 3$  was
generalized to allow for variability in $\nnu$. There a more general likelihood
function ${\cal L}(\eta, \nnu)$ was applied to the current best estimates of
the primordial \he4 and \li7 abundances.  Based on the analysis in \cite{ostsk},
we chose $\yp = 0.234 \pm 0.002 \rm (stat.) \pm 0.005 (syst.)$ as well as the
lower value $\yp = 0.230 \pm 0.003 \rm (stat.) \pm 0.005 (syst.)$ based on a
low metallicity subset of the data. Using these values of $\yp$ along with the
value (Li/H)$_p = (1.6 \pm 0.07) \times 10^{-10}$ from \cite{mol}, we found 
peak likelihood values $\eta_{10} = 1.8$ and $\nnu = 3.0$ with a 95\% CL range
of  $1.6\le\nnu\le4.0,
1.3\le\eta_{10}\le 5.0 $ for the higher \he4 value and similar results for
the lower one.  More recent data from Izotov and Thuan
\cite{iz2} seems to indicate a still higher value for $\yp$, and for this
reason as well as wishing to be independent of the ``{\em
current}" best estimate of the abundances, we derive our results for a
wide range of possible values for $Y_p$ and (Li/H)$_p$ which will account
for the possibility of stellar depletion for the latter \cite{pinn}.
Finally, in \cite{oth2}, we considered only the effect of the high D/H
value from quasar absorption systems. Since there was virtually no
overlap between the likelihood functions based on the low D/H value and
the other two elements, there was little point in using that value in our analysis. 
Since then, the low D/H value has been raised somewhat, and that together
with our present consideration of higher $\yp$ and (Li/H)$_p$ values
makes the exercise worth while.

In this paper, we follow the approach of \cite{oth2} -- \cite{fkot} in
constraining the theory 
 on the basis of the \he4 and \li7 data and to a lesser extent D/H, by constructing
a likelihood function ${\cal L}(\eta,\nnu)$. We discuss the current status of
the data in section 2, and indicate what range of values for the
primordial abundances we consider.   In section 3,  we display the likelihood 
functions we use. As this was discussed in more detail in \cite{oth2,fkot}, we
will be brief here.  Our results are given in section 4, and we draw  conclusions
in section 5.

\section{Observational Data}

Data pertinent to the primordial \he4 abundance is obtained from observations
of extragalactic HII regions.  These regions have low 
metallicities (as low as 1/50th solar), and thus are presumably more primitive than
similar regions in our own Galaxy.  The \he4 abundance used to extract a
primordial value spans roughly an order of magnitude in metallicity (e.g. O/H).
Furthermore, since there have been a considerable number of such systems observed
with metallicities significantly below solar, modeling plays a relatively
unimportant role in obtaining the primordial abundance of \he4 (see e.g.
\cite{fdo2}).

The \he4 data based on observations in \cite{p,iz} were discussed in
detail in \cite{ostsk}. There are over 70 such regions observed with
metallicities ranging  from about 2--30\% of solar metallicity. This data led
to the determination of a primordial \he4 abundance of 
$\yp = 0.234 \pm 0.002 \rm (stat.) \pm 0.005 (syst.)$
used in \cite{oth2}. That the statistical error is small is due to the large
number of regions observed and to the fact that the \he4 abundance in these
regions is found to be very well correlated to metallicity. 
In fact, as can be understood from the remarks which follow, the primordial
\he4 abundance is dominated by systematic rather than statistical
uncertainties.

The compilation in \cite{ostsk} included the data of \cite{iz}. Although
this data is found to be consistent with other data on a point by point
basis, taken alone, it would imply a somewhat higher primordial \he4
abundance.
Furthermore, the resulting value of $\yp$ depends on the method of data
analysis. When only
\he4 data is used to self-consistently determine the \he4 abundance (as
opposed to using other data such as oxygen and sulphur to determine the
parameters which characterize the HII region and are needed to convert an
observation of a
\he4 line strength into an abundance), a value of $\yp$ as high as $0.244 \pm
0.002$ can be found\footnote{We note that this method has been criticized as it
relies on some \he4 data which is particularly uncertain, and these
uncertainties have not been carried over into the error budget in the \he4
abundance \cite{ostsk}.}
\cite{iz}. 

The problem concerning \he4 has been accentuated recently with new data from
Izotov and Thuan \cite{iz2}.  The enlarged data set from \cite{p,iz2} was
considered in \cite{fdo2}.  The new resulting value for $\yp$ is 
\beq
\yp = 0.238 \pm 0.002 \rm (stat.) \pm 0.005 (syst.)
\label{eq:he4}
\eeq
The new data taken alone gives $\yp = 0.2444 \pm 0.0015$ 
when using the method based on a set of 5 helium recombination lines 
to determine all of the H II region
parameters.  By more conventional methods, the same data gives $\yp =
0.239 \pm 0.002$.  As one can see, the \he4 data is clearly dominated by
systematic uncertainties. 

There has been considerably less variability in the \li7 data over the last
several years. The \li7 abundance is determined by the observation of Li
in the atmospheres of old halo dwarf stars as a function of metallicity (in
practice, the Fe abundance). The abundance used in \cite{oth2} from the work in
\cite{mol} continues to lead to the best estimate of the \li7 abundance in the
so called Spite plateau
\beq
y_7 \equiv \frac{\li7}{\rm H} = (1.6 \pm 0.07) \times 10^{-10}
\label{eq:li7}
\eeq
where the error is statistical, again due to the large number of stars observed. 
If we employ the basic  chemical evolution conclusion that metals
increase linearly with time, we may infer this value to be indicative of the
primordial Li abundance.

In \cite{oth2}, we noted that there  are considerable systematic uncertainties
in the plateau abundance. It is often questioned as to whether the Pop II
stars  have preserved their initial abundance of Li.  
While the detection of the more fragile isotope \li6 in two of
these stars may argue against a strong depletion of \li7
\cite{sfosw,pinn}, it is difficult to exclude depletion of the order of a
factor of two. Therefore it seems appropriate to allow for a wider
range in \li7 abundances in our likelihood analysis than was done in
\cite{oth2}.

There has been some, albeit small, change in the D/H data from quasar absorption
systems.  Although the re-observation of the high D/H in \cite{rh1} has been
withdrawn,  the original measurements \cite{quas1} of this object still
stand at the high value. More recently, a different system at the relatively
low redshift of $z = 0.7$ was observed to yield a similar high value
\cite{quas3}
\beq
y_2 \equiv {\rm D/H} = (2 \pm 0.5) \times 10^{-4}.
\label{dhigh}
\eeq
The low
values of D/H in other such systems reported in \cite{quas2} have since been
refined to show slightly higher D/H values \cite{quas4}
\beq
y_2 \equiv {\rm D/H} = (3.4 \pm 0.3) \times 10^{-5}.
\label{dlow}
\eeq 
Though this value is still significantly lower than the high D/H value
quoted above, the low value is now high enough that it contains
sufficient overlap with the ranges of the other light elements considered
to warrant its inclusion in our analysis.

\section{Likelihood Functions}

Monte Carlo and likelihood analyses have been discussed at great length in the
context of BBN \cite{kr,skm,kk1,kk2,hata1,hata2,fo,fkot,oth2}.
Since our likelihood analysis follows that of \cite{fkot} and \cite{oth2},
we will be very brief here.
The likelihood function for \he4, $L_4(\nnu, \eta)$ is determined from a
convolution of a theory function
\beq
L_{\rm 4,Theory}(Y, \nnu, \eta) =
    {1\over\sqrt{2\pi}\sigma_{Y}(\nnu,\eta)}
    \exp{\left({-(Y-\yp(\nnu,\eta))^{2}\over2\sigma_{Y}^{2}(\nnu,\eta)}\right)}
\eeq
(where $\yp(\nnu,\eta)$ and $\sigma_{Y}(\nnu,\eta)$ represent the results 
of the theoretical calculation) and an observational function
\beq
L_{\rm 4,Obs}(Y) =
    {1\over\sqrt{2\pi}\sigma_{Y0}}
    \exp{\left({-(Y-Y_0)^{2}\over2\sigma_{Y0}^{2}}\right)}
\eeq
where $Y_0$ and $\sigma_{Y0}$ characterize the observed 
distribution and are taken from Eqs. (\ref{eq:he4}) and (\ref{eq:li7}).
 The full likelihood function for \he4 
is then given by
\beq
L_{4}(\nnu, \eta) =
    \int dY\, L_{\rm 4,Theory}(Y, \nnu, \eta) L_{\rm 4, Obs}(Y)
\eeq
which can be integrated (assuming Gaussian errors as we have done) to give
\beq
L_{4}(\nnu, \eta) =
{1\over\sqrt{2\pi(\sigma_Y^2(\nnu,\eta)+\sigma_{Y0}^2)}}
\exp\left({-(\yp(\nnu,\eta)-Y_0)^2\over 
2(\sigma_Y^2(\nnu,\eta)+\sigma_{Y0}^2)}\right)
\eeq

The likelihood functions for \li7 and D are constructed in a similar
manner. The quantities of interest in constraining the $\nnu$---$\eta$ plane 
are the combined likelihood functions
\beq
L_{47} = L_4\times L_7
\eeq
and
\beq
L_{247} = L_{2}\times L_{47}.
\eeq
Contours of constant $L_{47}$ (or $L_{247}$ when we include D in the analysis)
represent equally likely points in the 
$\nnu$--$\eta$ plane.  Calculating the contour containing 95\% of 
the volume under the $L_{47}$ surface gives us the 95\% likelihood 
region. From these contours we can then read off ranges of $\nnu$ and $\eta$.  

\section{Results}

Using the abundances in eqs (\ref{eq:he4},\ref{eq:li7}) and adding
the  systematic errors to the statistical errors in quadrature we have
a maximum likelihood distribution, $L_{47}$, which is shown in
Figure 1a. This is very similar to our previous result based on
the slightly lower value of $\yp$.  As one can see, $L_{47}$ is double peaked. 
This is due to the minimum in the predicted lithium abundance as a function of
$\eta$, as was discussed earlier. 
We also show in Figures 1b and 1c, the resulting likelihood distributions,
$L_{247}$, when
the high and low D/H values from Eqs. (\ref{dhigh}) and (\ref{dlow}) are
included.

The peaks of the distribution as well as the allowed ranges of
$\eta$ and
$\nnu$ are more easily discerned in the  contour plots of Figure 2 which shows
the 50\%,  68\% and 95\% confidence level contours in $L_{47}$ and $L_{247}$.  
The crosses show the location of the 
peaks of the likelihood functions.  Note that
$L_{47}$ peaks at $\nnu=3.2$, (up slightly from the case with $\yp =
.234$) and $\eta_{10}=1.85$. The second peak of $L_{47}$ occurs at 
$\nnu=2.6$, $\eta_{10}=3.6$.  The 95\% confidence level allows the
following ranges in $\eta$ and $\nnu$
\begin{eqnarray}
1.7\le\nnu\le4.3 \nonumber \\
%1.3\le~\eta_{10}~\le 5.0 
1.4\le\eta_{10}\le4.9
\end{eqnarray}
These results differ only slight from those in \cite{oth2}.

Since $L_{2}$ picks out a small range of values 
of $\eta$, largely independent of $\nnu$, its effect on $L_{247}$ is 
to eliminate one of the two peaks in $L_{47}$.
With the high D/H value,  $L_{247}$ 
peaks at the slightly higher value $\nnu=3.3$, 
$\eta_{10}=1.85$. In this case the 95\% contour gives the ranges
\begin{eqnarray}
2.2\le\nnu\le4.4 \nonumber \\
%1.4\le~\eta_{10}~\le 2.6 
1.4\le\eta_{10}\le 2.4 
\end{eqnarray}
(Strictly speaking, $\eta_{10}$ can also be in the range
3.2---3.5, with $2.5\la\nnu\la2.9$ as can be seen by the 95\% contour in
Figure 2a. However this ``peak" is almost completely invisible in Figure
1b.) The 95\% CL ranges in $\nnu$ for both $L_{47}$ and $L_{247}$
include values below the canonical value $\nnu = 3$. Since one could
argue that $\nnu \ge 3$, we could use this condition as a Bayesian
prior.  This was done in \cite{osb} and in the present context in
\cite{oth2}.  In the latter, the effect on the limit to $\nnu$ was
minor, and we do not repeat this analysis here.

In the case of low D/H,
$L_{2}$ picks out a smaller value of
$\nnu = 2.4$ and a larger value of $\eta = 4.55$. 
The 95\% CL upper limit is now $\nnu < 3.2$, and the range for 
$\eta$ is $ 3.9 < \eta_{10} < 5.4$.  It is important to stress that with
the increase in the determined value of D/H \cite{quas4} in the low D/H
systems, these abundances are now consistent with the standard model
value of $\nnu = 3$ at the 2 $\sigma$ level.

Although we feel that the above set of values represents the {\em current} best choices
for the observational parameters, our real goal in this paper is to generalize these
results for a wide range of possible primordial abundances.
To begin with, we will fix (Li/H)$_p$ from Eq. (\ref{eq:li7}), and allow 
$\yp$ to vary from 0.225 -- 0.250. In Figure 3, the positions of the two peaks of the
likelihood function, $L_{47}$, are shown as functions of $\yp$.  The low-$\eta$ peak is shown
by the dashed curve, while the high-$\eta$ peak is shown as dotted. The preferred
value of $\nnu = 3$, corresponds to a peak of the likelihood function either at $\yp
= 0.234$ at low  $\eta_{10} = 1.8$ or at $\yp = 0.243$ at $\eta_{10} = 3.6$
(very close to the value of $\yp$ quoted in \cite{iz2}). Since the peaks
of the likelihood function are of comparable height, no useful
statistical information can be extracted concerning the relative
likelihood of the two peaks.  The 95\% CL upper limit to
$\nnu$ as a function of $\yp$ is shown by the solid curve, and over the range in $\yp$
considered varies from 3.3 -- 5.3.
The fact that the peak value of $\nnu$ (and its upper limit) increases with $\yp$ is
easy to understand. The BBN production of \he4 increases with increasing $\nnu$. 
Thus for fixed Li/H, or fixed $\eta$, 
raising $\yp$ must be compensated for by raising $\nnu$ in order to avoid moving 
the peak likelihood to higher values of $\eta$ and therefore off of the \li7 peak.  

In Figure 4, we show the corresponding results with (Li/H)$_p = 4.1 \times 10^{-10}$.
In this case,  we must assume that lithium was depleted by a factor of $\sim
2.5$ or 0.4 dex, corresponding to the upper limit derived in \cite{pinn}.
The effect of assuming a higher value for the primordial abundance of Li/H is that
the two peaks in the likelihood function are split apart.  
Now the value of $\nnu = 3$ occurs at $\yp = 0.227$ at $\eta_{10} = 1.1$ (a very low
value) and at $\yp = 0.248$ and $\eta_{10} = 5.7$.  The 95\% CL upper
limit   on $\nnu$ in this case
can even extend up to 6 at $\yp = 0.250$. 
In Figure 5, we show a compilation of the 95\% CL upper limits to $\nnu$
for different values of (Li/H)$_p = 1.6, 2.0, 2.6, 3.2, {\rm and} ~4.1
\times 10^{-10}$. The upper limit to $\nnu$ can be approximated by a fit
to our results which can be expressed as
\beq
\nnu \la  80 \yp + 2.5 \times 10^9 {\rm (Li/H)}_p - 15.15
\eeq

Finally we turn to the cases when D/H from quasar absorption systems are
also considered in the analysis. For the high D/H given in Eq.
(\ref{dhigh}), though there is formally still a high-$\eta$ peak, the
value of the likelihood function
$L_{247}$ there is so low that it barely falls within the 95\% CL equal
likelihood contour (see Figures 1b and 2a). Therefore
we will ignore it here.  In Figure 6, we show the peak value of $\nnu$
and its upper limit for the two cases of (Li/H)$_p = 1.6$ and $4.1 \times
10^{-10}$. These results differ only slightly from those shown in Figures
3 and 4.   We note however, that overall the two values of Li/H do not
give an equally good fit. For fixed D/H, the high value prefers a value
of $\eta_{10} \simeq 1.8$ coinciding with the position of the low-$\eta$
peak for (Li/H)$_p = 1.6 \times 10^{-10}$.  At higher Li/H, the
low-$\eta$ peak shifts to lower $\eta$ diminishing the overlap with D/H.
In fact at (Li/H)$_p \ga 3.8 \times 10^{-10}$, the likelihood function
$L_{247}$ takes peak values which would lie outside the 95\% CL contour
of the case (Li/H)$_p = 1.6 \times 10^{-10}$. The relative values of the
likelihood function $L_{247}$, on the low-$\eta$ peak, for the five values of Li/H
considered are shown in Figure 7. Contrary to our inability to statistically
distinguish between the two peaks of $L_{47}$, the large variability in the values
of
$L_{247}$ shown in Figure 7 are statistically relevant. Thus,
as claimed in \cite{fkot}, if the high D/H could be confirmed, one could
set a strong limit on the amount of \li7 depletion in halo dwarf stars.

Since the low D/H value has come up somewhat, and since here we are
considering the possibility for higher values of $\yp$ and (Li/H)$_p$,
the statistical treatment of the low D/H case is warranted.  In Figure 8,
we show the peak value and 95\% CL upper limit from $L_{247}$ when the
low value of D/H is used from Eq. (\ref{dlow}) with (Li/H)$_p = 1.6
\times 10^{-10}$. The results are not significantly different in this
case for the other choices of (Li/H)$_p$. In order to obtain $\nnu = 3$,
one needs to go to \he4 abundances as high as $\yp = 0.247$ with respect
to the peak of the likelihood function. However, for $Y_p > 0.234$, the
revised low value of D/H is compatible with \he4 and \li7 at the 95\%
CL.  The likelihood functions $L_{247}$ are shown in Figure 9 for
completeness.

\section{Conclusions}

We have generalized the full two-dimensional (in $\eta$ and $\nnu$)
likelihood analysis based on big bang nucleosynthesis for a wide range of
possible primordial abundances of \he4 and \li7. Allowing for full freedom
in both the baryon-to-photon ratio,
$\eta$, and the number of light particle degrees of freedom as
characterized by the number   of light, stable neutrinos, $\nnu$, we have
updated the allowed range in $\eta$ and $\nnu$ based the higher value of
$Y_p = 0.238 \pm 0.002 \pm 0.005$ from \cite{fdo2} which includes the
recent data in \cite{iz2}. The likelihood analysis based on \he4 and
\li7 yields the 95\% CL upper limits: $\nnu \le 4.3$ and $\eta_{10} \le
4.9$.  The result for $\nnu$ is only slightly altered, $\nnu \le 4.4$,
when the high values of D/H observed in certain  quasar absorption systems
\cite{quas1,quas3} are included in the analysis. In this case, the upper limit to
$\eta_{10}$ is lowered to 2.4.
 Since the low values of D/H have been revised upward somewhat
\cite{quas2}, they are now consistent with \he4 and \li7 and $\nnu = 3$
at the 95\% CL.
We have also shown how our results for the upper limit to $\nnu$ depend
on the specific choice for the primordial abundance of \he4 and \li7.
If we assume that the observational determination of \li7 in halo stars
is a true indicator the primordial abundance of \li7, then the upper
limit to $\nnu$ varies from 3.3 -- 5.3 for $\yp$ in the range 0.225 --
0.250.  If on the other hand, \li7 is depleted in halo stars  by as much
as a factor of 2.5, then the upper limit to $\nnu$ could extend up to 6
at $\yp = 0.250$.

\bigskip

{\bf Acknowledgments}
We note that this work was begun in collaboration with David Schramm.
This work was supported in part by 
DOE grant DE-FG02-94ER40823 at Minnesota.

\newpage

\begin{figure}[tbp]
	\centering
	\epsfysize=5in \epsfbox{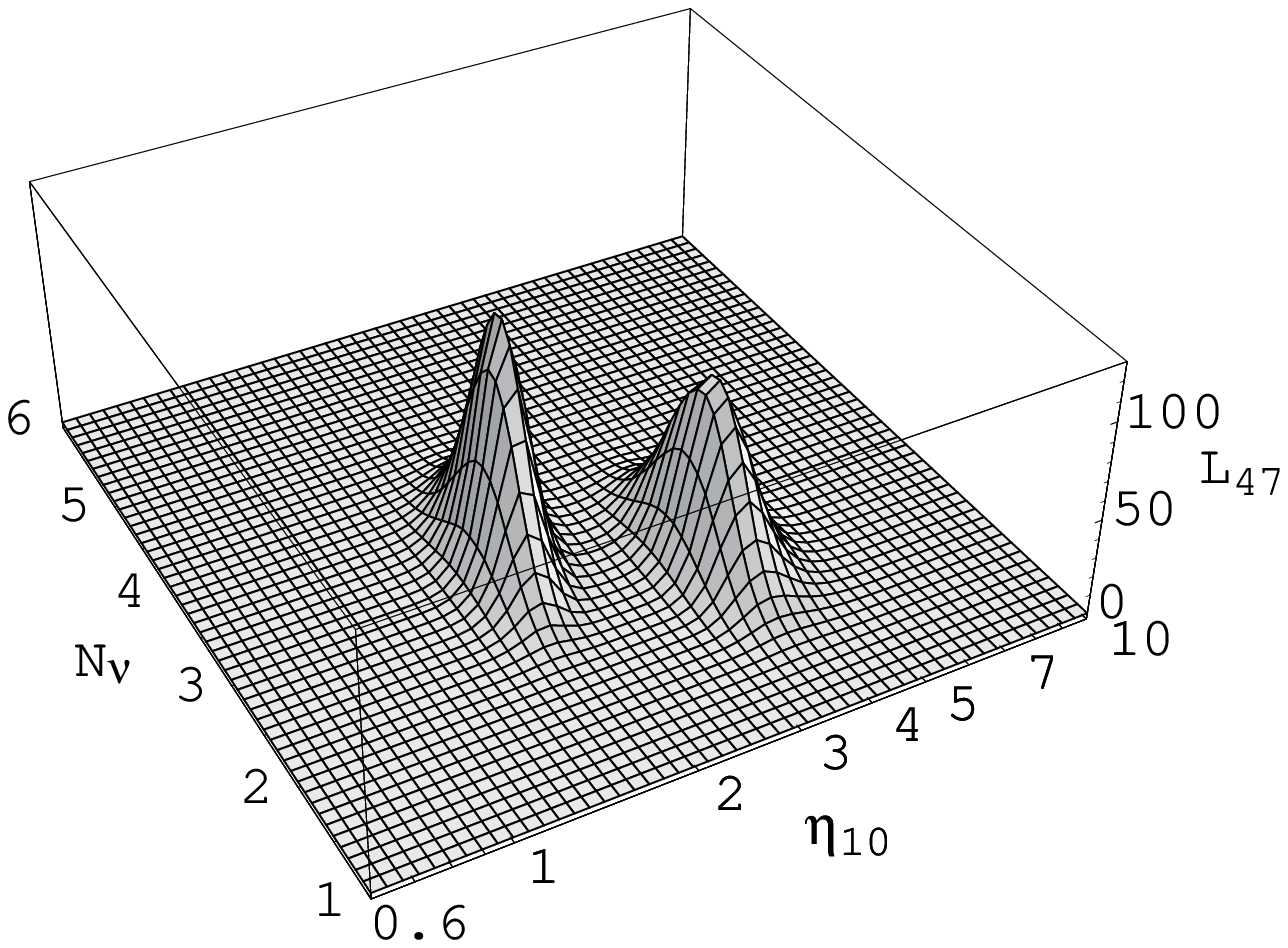}
	\caption{(a) $L_{47}(N_\nu, \eta)$ for observed abundances given by
                 eqs. (\protect\ref{eq:he4} and \protect\ref{eq:li7}).}
	\label{fig1a}
\end{figure}
\setcounter{figure}{0}
\begin{figure}[tbp]
	\centering
	\epsfysize=5in \epsfbox{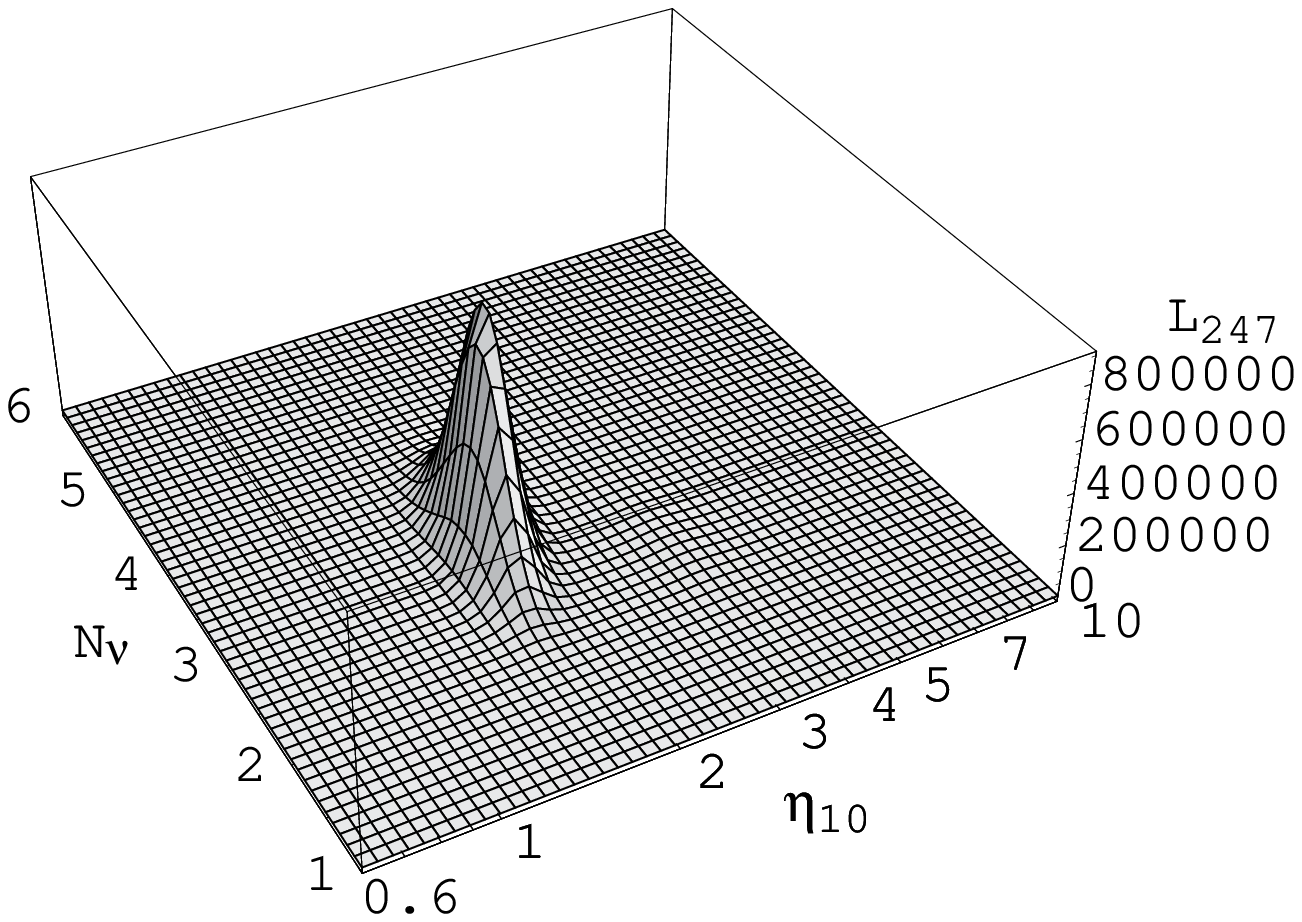}
	\caption{(b) $L_{247}(N_\nu, \eta)$ for observed abundances given by
                 eqs. (\protect\ref{eq:he4}, \protect\ref{eq:li7}, and
\protect\ref{dhigh}).}
	\label{fig1b}
\end{figure}
\setcounter{figure}{0}
\begin{figure}[tbp]
	\centering
	\epsfysize=5in \epsfbox{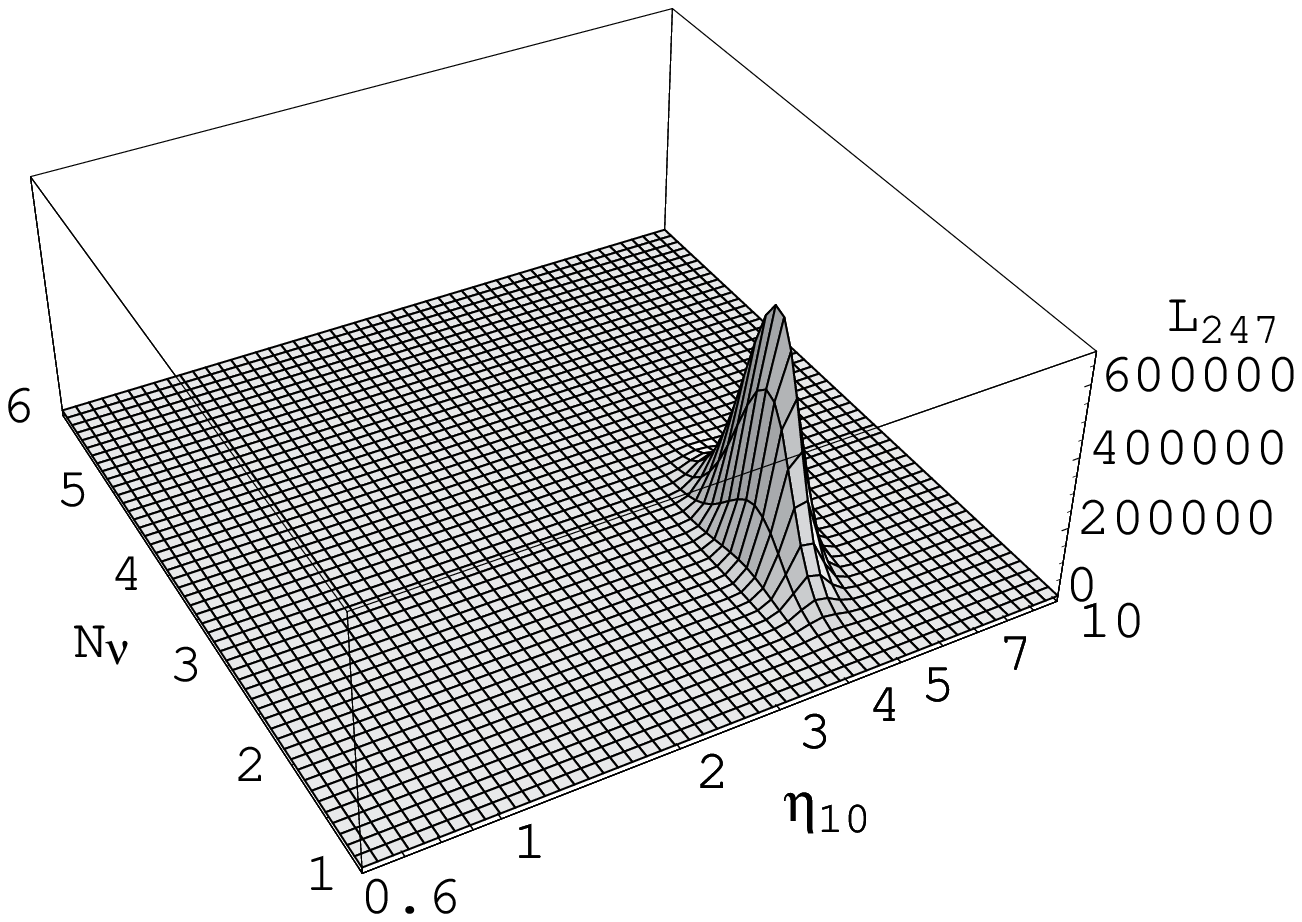}
	\caption{(c) $L_{247}(N_\nu, \eta)$ for observed abundances given by
                 eqs. (\protect\ref{eq:he4}, \protect\ref{eq:li7}, and
\protect\ref{dlow}).}
	\label{fig1c}
\end{figure}

\begin{figure}[tbp]
	\centering
	\epsfysize=7in \epsfbox{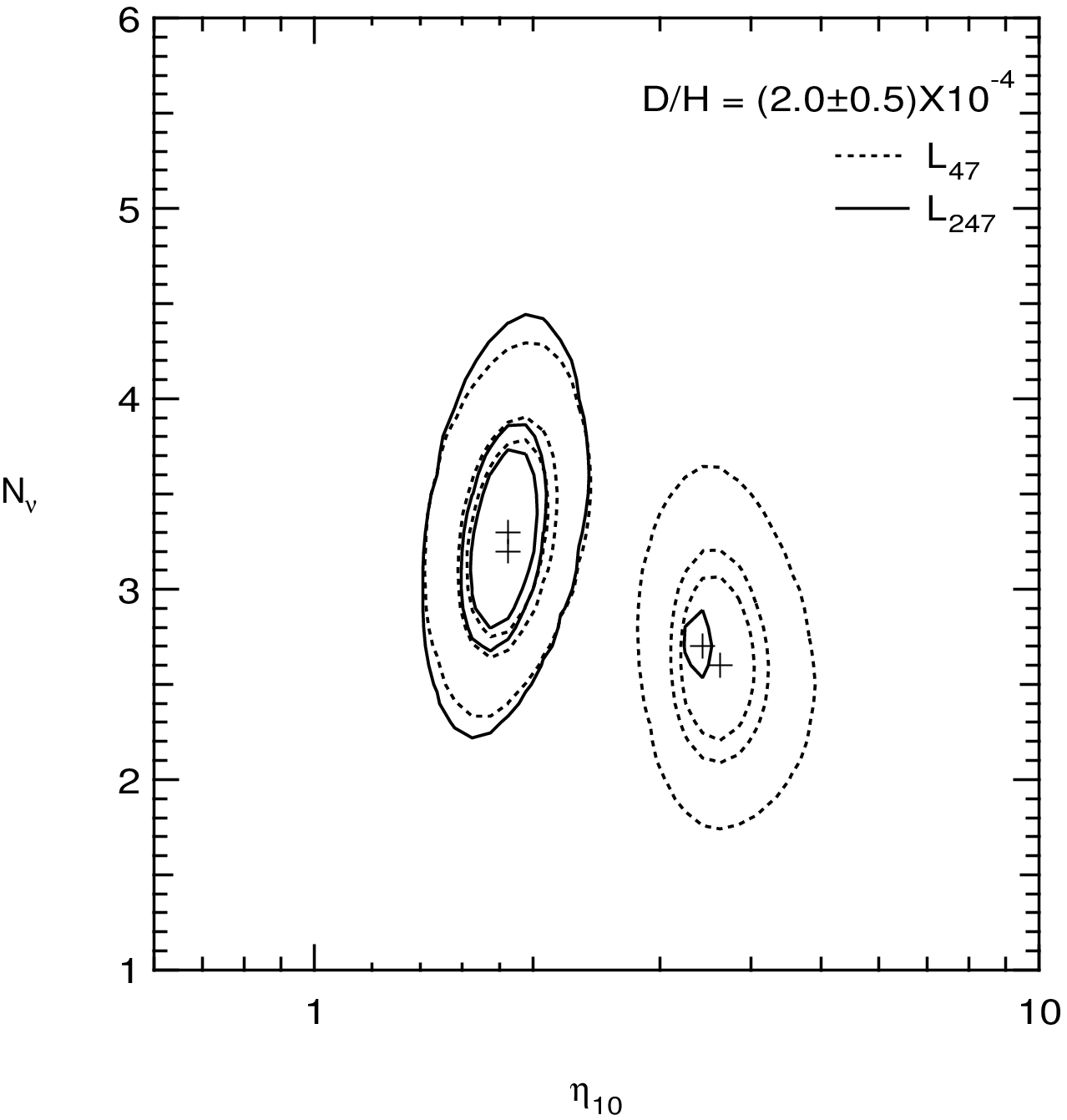}
	\caption{(a) 50\%, 68\% \& 95\% C.L. contours of $L_{47}$ and
                 $L_{247}$ where observed abundances are given by
                 eqs. (\protect\ref{eq:he4}, \protect\ref{eq:li7}, and
\protect\ref{dhigh}).}
	\label{fig2a}
\end{figure}
\setcounter{figure}{1}
\begin{figure}[tbp]
	\centering
	\epsfysize=7in \epsfbox{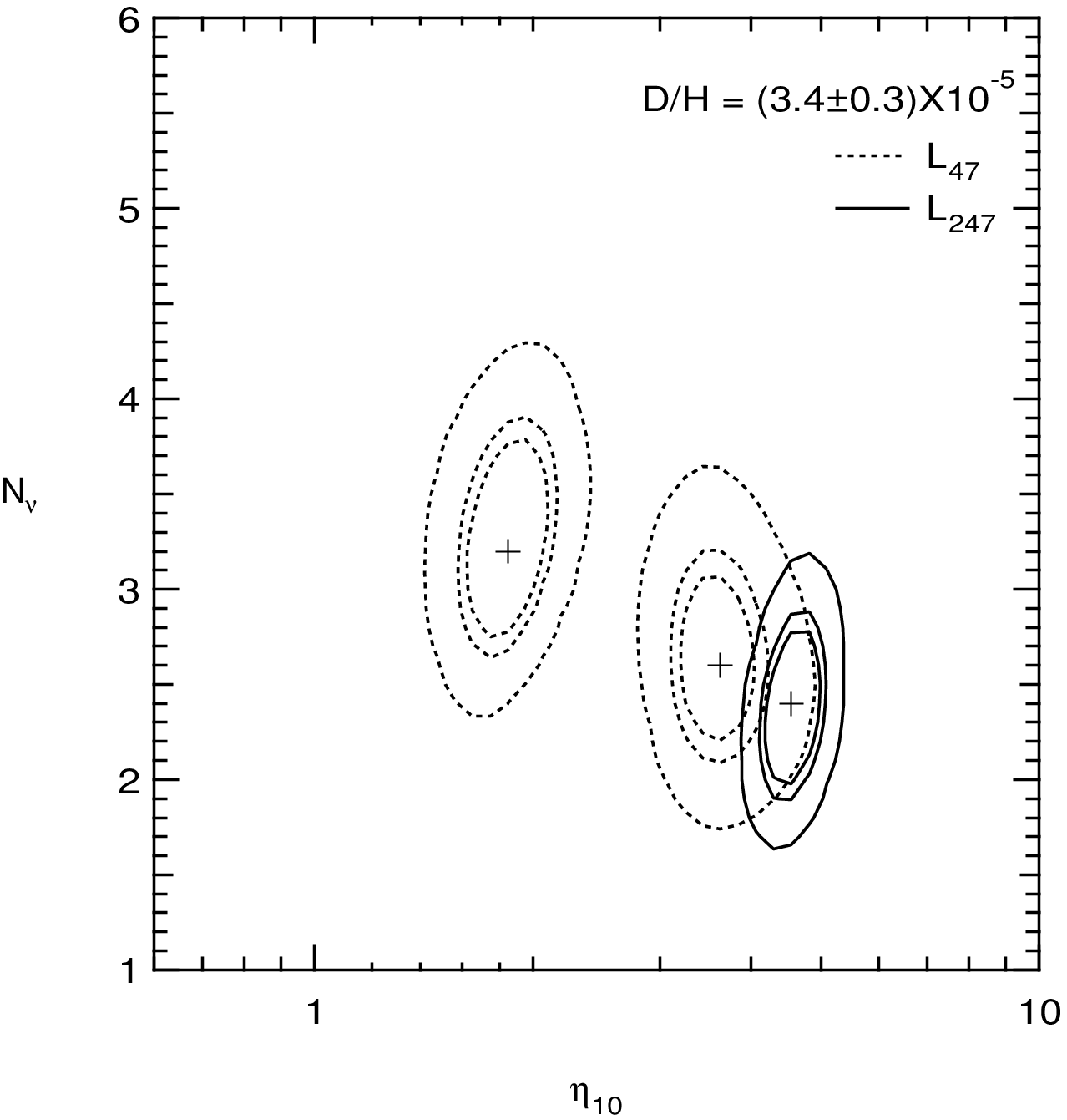}
	\caption{(b) 50\%, 68\% \& 95\% C.L. contours of $L_{47}$ and
                 $L_{247}$ where observed abundances are given by
                 eqs. (\protect\ref{eq:he4}, \protect\ref{eq:li7}, and
\protect\ref{dlow}).}
	\label{fig2b}
\end{figure}

\begin{figure}[tbp]
	\centering
	\epsfysize=4in \epsfbox{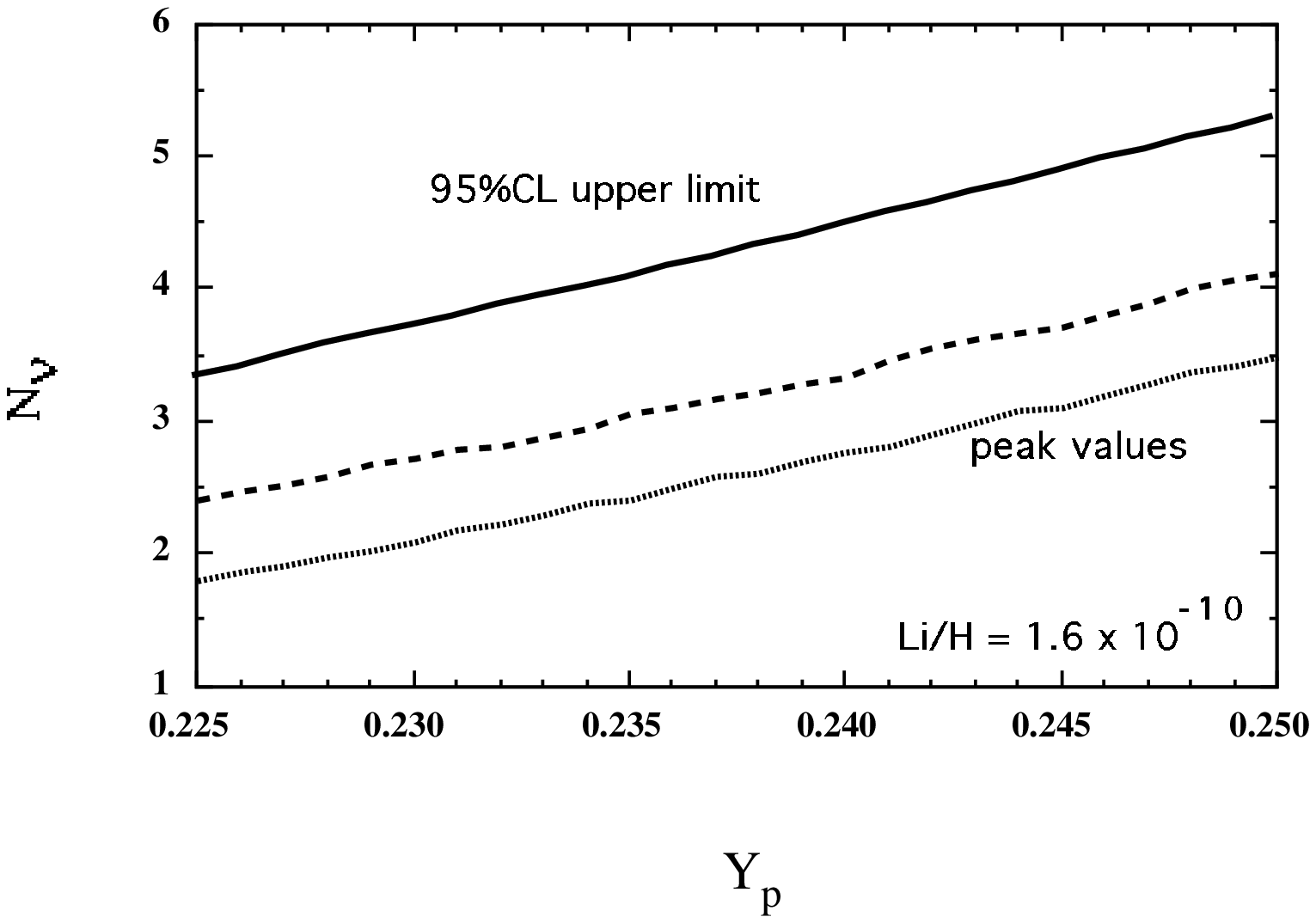}
	\caption{The position of the value of $\nnu$ along the low-$\eta$ peak (dashed) and
high-$\eta$ peak (dotted) of the likelihood function $L_{47}$ as function of $\yp$. 
The solid  curve shows the 95\% CL upper limit to $\nnu$ as a function of $\yp$.
The value of (Li/H)$_p = 1.6 \times 10^{-10}$ has been fixed.}
	\label{fig3}
\end{figure}

\begin{figure}[tbp]
	\centering
	\epsfysize=4in \epsfbox{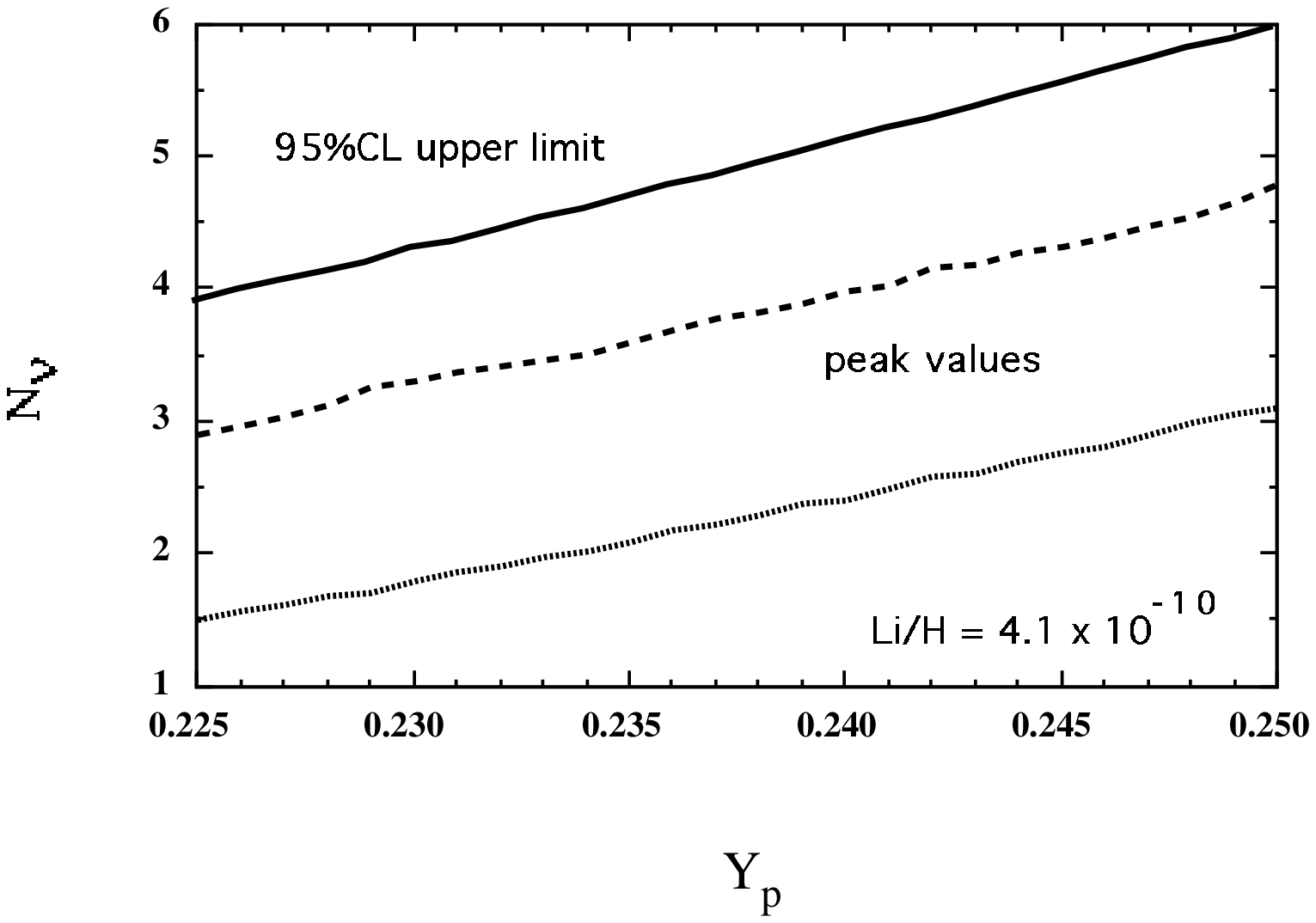}
	\caption{As in Figure \protect\ref{fig3} with  (Li/H)$_p =4.1 \times 10^{-10}$.}
	\label{fig4}
\end{figure}

\begin{figure}[tbp]
	\centering
	\epsfysize=4in \epsfbox{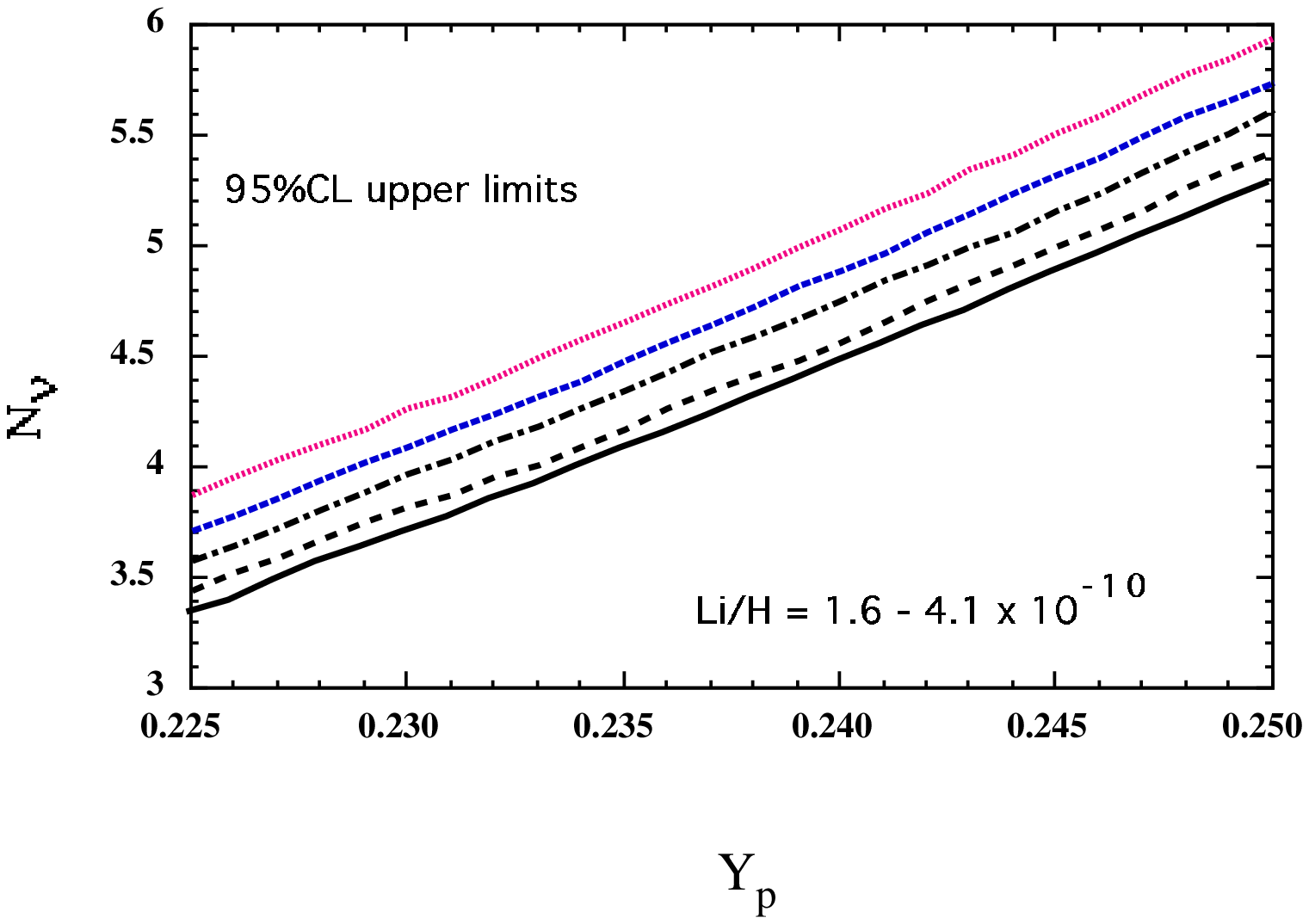}
	\caption{Summary of the upper limits to $\nnu$ for (Li/H)$_p = 1.6, 2.0, 2.6, 3.2,
{\rm and}\  4.1 \times 10^{-10}$ as a function of $\yp$. The lowest curve
corresponds to (Li/H)$_p = 1.6 \times 10^{-10}$ and the limits on $\nnu$ increase
with (Li/H)$_p$.}
	\label{fig5}
\end{figure}

\begin{figure}[tbp]
	\centering
	\epsfysize=4in \epsfbox{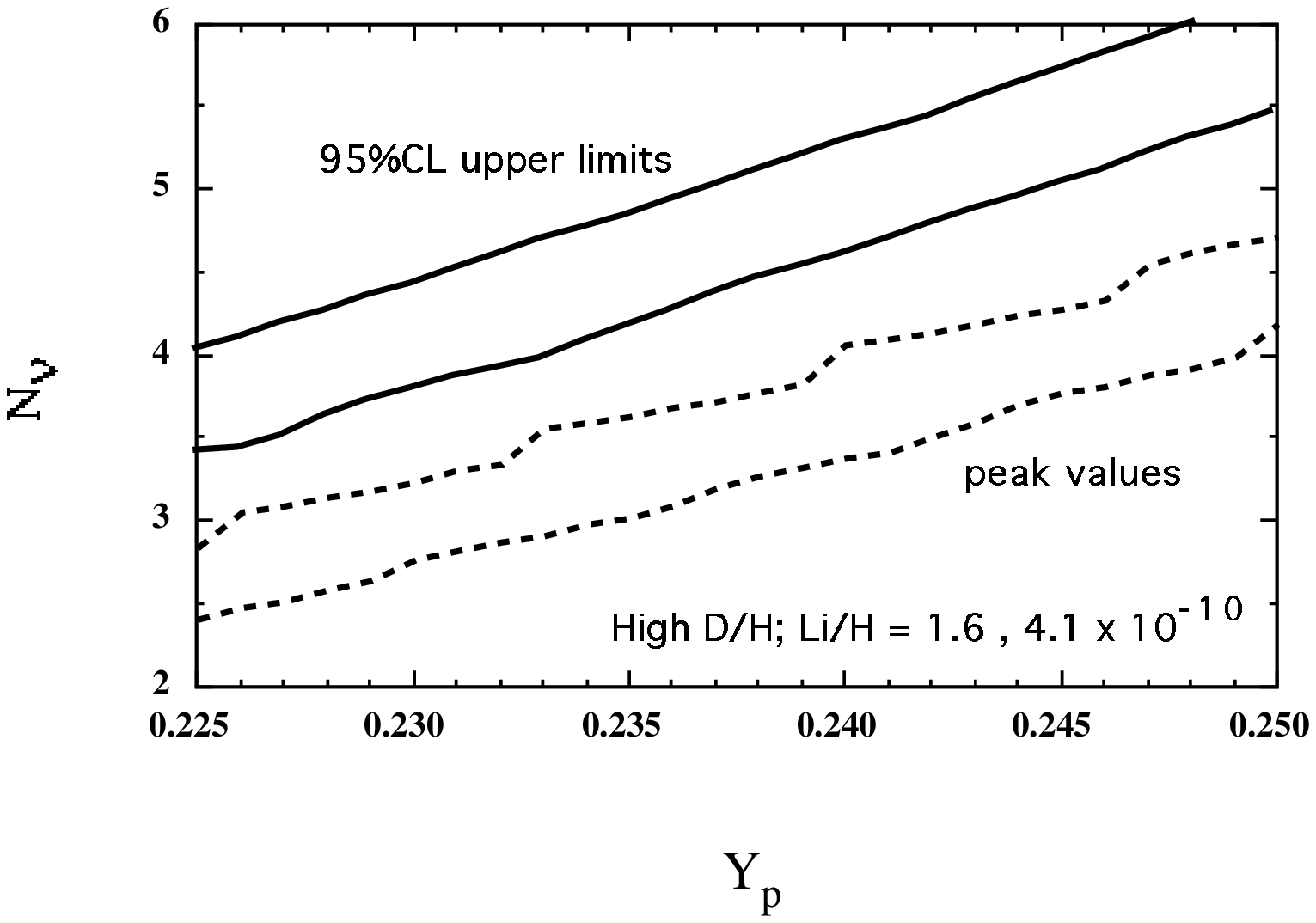}
	\caption{As in  Figures \protect\ref{fig3} and \protect\ref{fig4} based on the
likelihood function $L_{247}$ which includes high D/H from quasar absorption
systems.}
	\label{fig6}
\end{figure}

\begin{figure}[tbp]
	\centering
	\epsfysize=4in \epsfbox{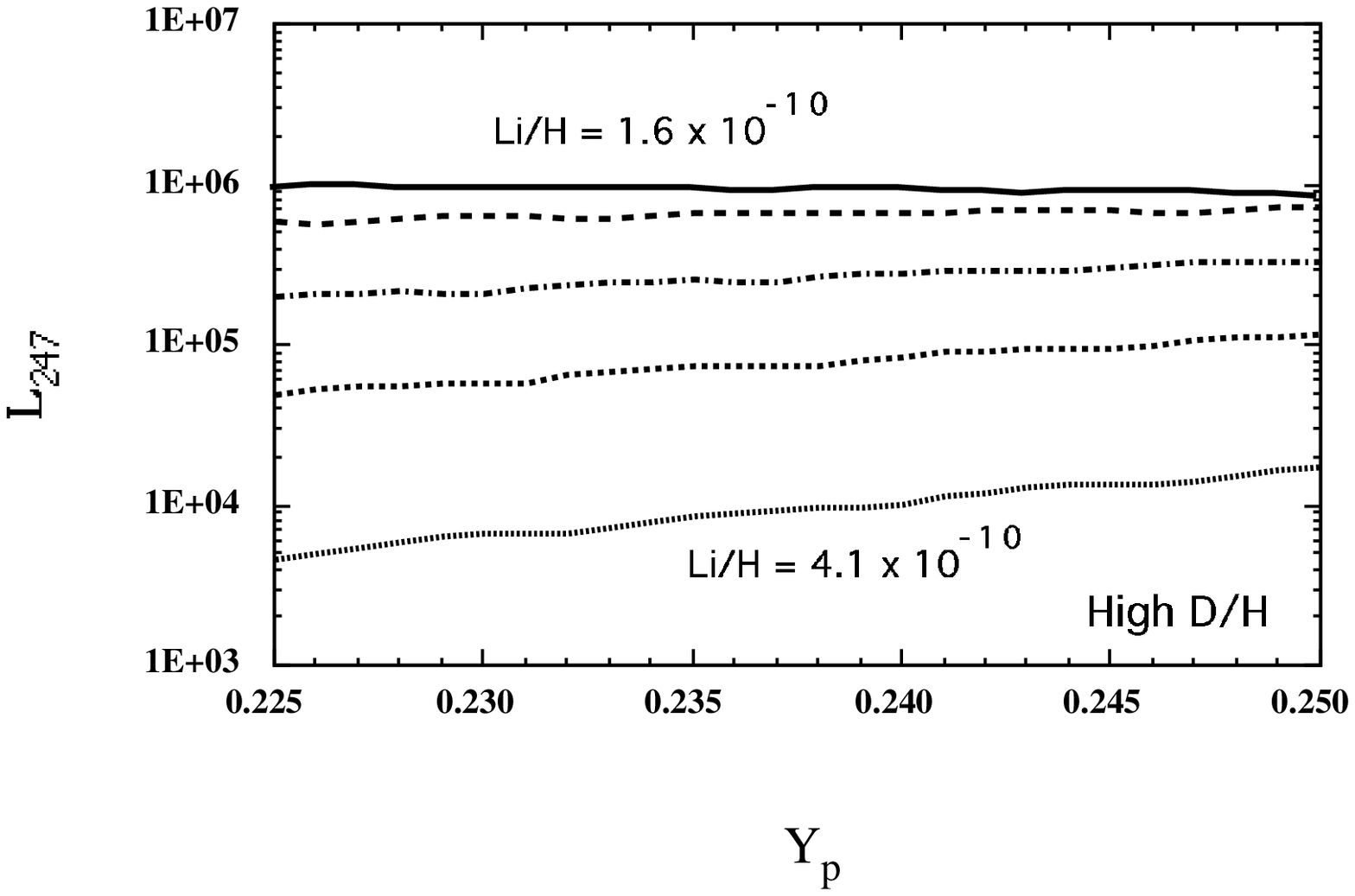}
	\caption{Relative values of the likelihood function $L_{247}$, on the low-$\eta$
peak, for the five choices of (Li/H)$_p$ in Figure \protect\ref{fig5}.}
	\label{fig7}
\end{figure}

\begin{figure}[tbp]
	\centering
	\epsfysize=4in \epsfbox{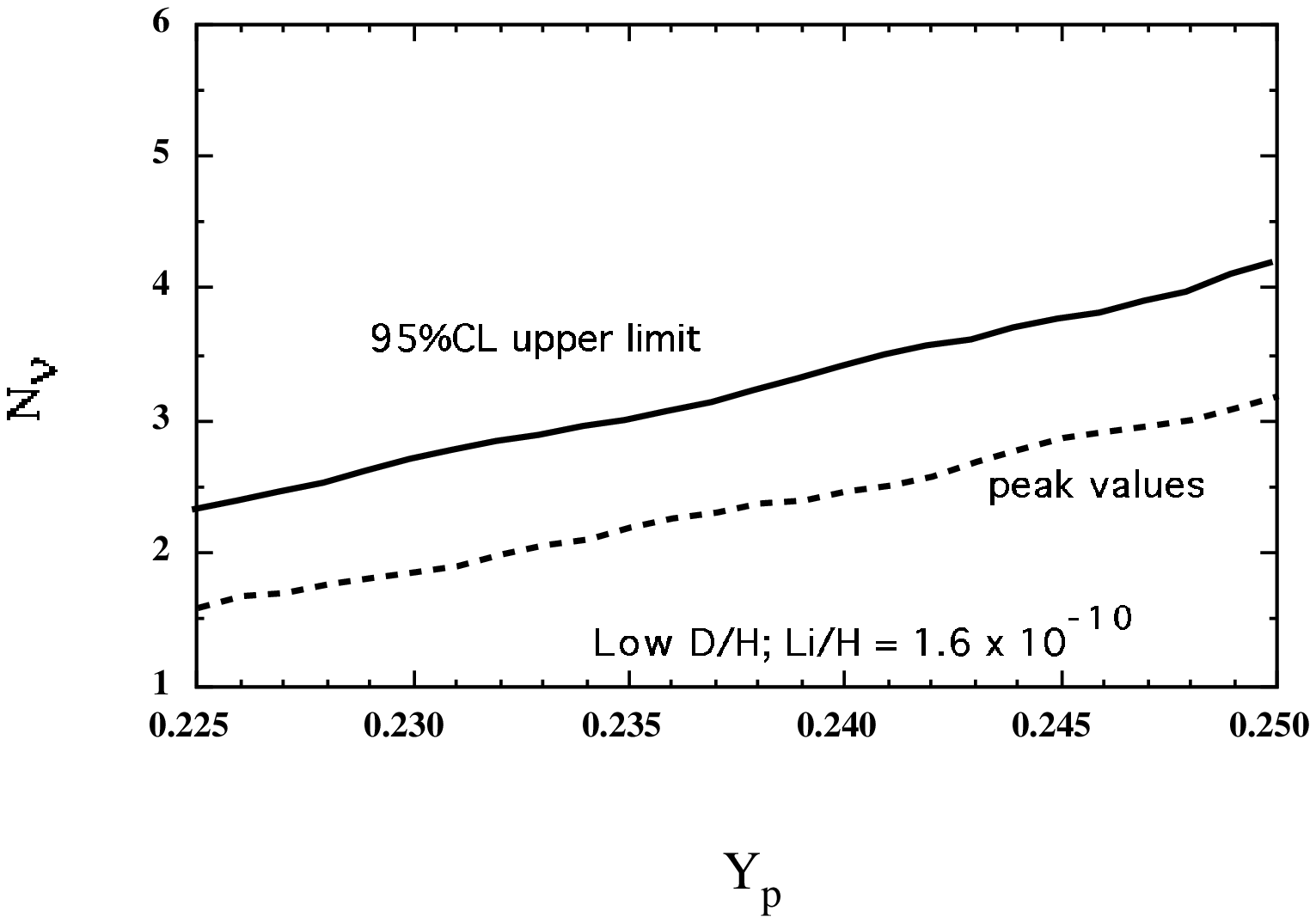}
	\caption{As in Figure \protect\ref{fig6} for low D/H from quasar absorption
systems.}
	\label{fig8}
\end{figure}

\begin{figure}[tbp]
	\centering
	\epsfysize=4in \epsfbox{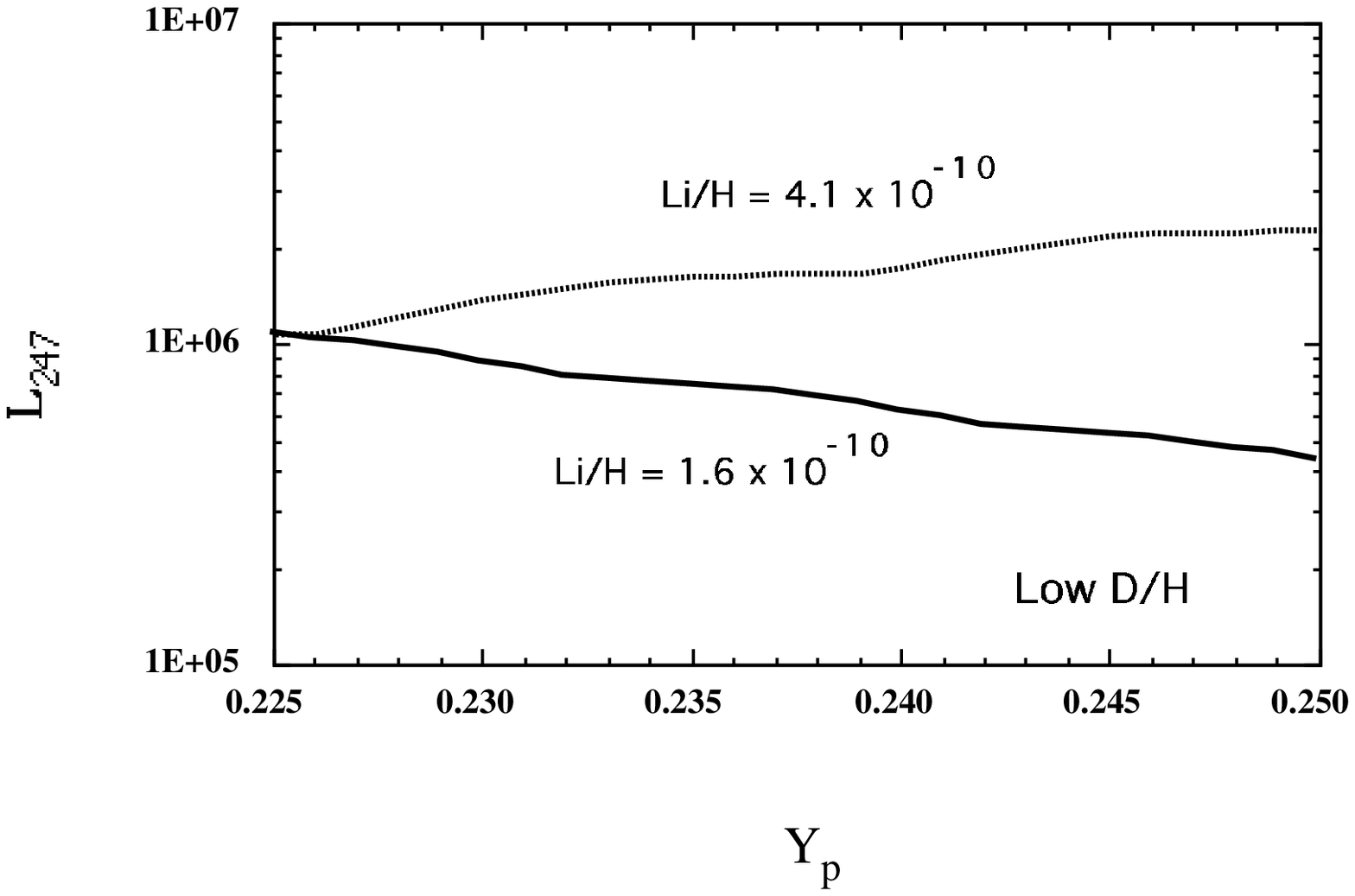}
	\caption{As in Figure \protect\ref{fig7} for low D/H from quasar absorption
systems.}
	\label{fig9}
\end{figure}


\begin{thebibliography}{99}

\bibitem{ssg} G. Steigman, D.N. Schramm, and J. Gunn,
  Phys. Lett. {\bf B66} (1977) 202.

\bibitem{oth2} K.A. Olive and D. Thomas, AstroPart. Phys. {\bf 7} (1997) 
27.

\bibitem{fo} B.D. Fields, and K.A. Olive, Phys. Lett. {\bf B368} (1996)
103.

\bibitem{fkot} B.D. Fields, K. Kainulainen, K.A. Olive, and D. Thomas,
 New Astr. {\bf 1} (1996) 77.

\bibitem{cst2} C.J. Copi, D.N. Schramm, and M.S. Turner,
Phys. Rev. {\bf D55} (1997) 3389; N. Hata, G. Steigman, S.
Bludman and P. Langacker, Phys. Rev. {\bf D55} (1997) 540.


\bibitem{ostsk} K.A. Olive, E. Skillman, and G. Steigman, Ap.J. {\bf
483} (1997) 788. 

\bibitem{mol} P. Molaro, F. Primas, and P. Bonifacio, A.A. {\bf 295 }
(1995) L47; P. Bonifacio and P. Molaro, MNRAS, {\bf 285} (1997) 847.

\bibitem{iz} Y.I. Izotov, T.X. Thuan, and V.A. Lipovetsky,
Ap.J. {\bf 435} (1994) 647; Ap.J.S. {\bf 108} (1997) 1.

\bibitem{pinn} M.H. Pinsonneault, T.P. Walker, G. Steigman, and V.K.
Naranyanan, Ap.J. (1998) submitted, astro-ph/9803073.

\bibitem{orstv} K.A. Olive, R.T. Rood, D.N. Schramm, J.W. Truran, and E.
Vangioni-Flam,  Ap.J. {\bf 444} (1995) 680;
D. Galli, F. Palla, F. Ferrini, and U. Penco,
Ap.J. {\bf 443} (1995) 536; D. Dearborn, G. Steigman, and M. Tosi,  Ap.J.
{\bf 465} (1996) 887; S.T. Scully, M. Cass\'{e}, K.A. Olive,
D.N. Schramm, J. Truran, and E. Vangioni-Flam, Ap.J. {\bf 462} (1996) 960;
K.A. Olive, S.T. Scully, D.N. Schramm, and J. Truran, Ap.J. {\bf 479} 
(1996) 752.

\bibitem{scov} S. Scully, M. Cass\'{e}, K.A. Olive, E. Vangioni-Flam,
Ap. J. {\bf 476} (1997) 521.

\bibitem{cce} S.J. Lilly, O. Le Fevre, F. Hammer, and D. Crampton,  
Ap.J. {\bf 460} (1996) L1; P. Madau, H.C. Ferguson, M.E.  Dickenson,
M. Giavalisco, C.C. Steidel, and A. Fruchter,  MNRAS {\bf 283} (1996)
1388; A.J. Connolly, A.S. Szalay, M. Dickenson, M.U. SubbaRao, and R.J.
Brunner, Ap.J. {\bf 486} (1997) L11; M.J. Sawicki, H. Lin, and H.K.C.
Yee, A.J. {\bf 113} (1997) 1.

\bibitem{cova} M. Cass\'e, K.A. Olive, E. Vangioni-Flam, and J. Audouze,
New Astronomy {\bf 3} (1998) 259.

\bibitem{quas1} R.F. Carswell, M. Rauch, R.J. Weymann, A.J. Cooke, and
J.K. Webb,  MNRAS {\bf 268} (1994) L1; A. Songaila, L.L. Cowie,  
C. Hogan, and M. Rugers,  Nature {\bf 368} (1994) 599.

\bibitem{quas2} D. Tytler, X.-M. Fan, and S. Burles,  Nature {\bf 381}
 (1996) 207; S. Burles and D. Tytler,  Ap.J. {\bf 460} (1996) 584.

\bibitem{quas3} J.K. Webb, R.F. Carswell, K.M. Lanzetta, R. Ferlet, M.
Lemoine, A. Vidal-Madjar, and D.V. Bowen,  Nature {\bf 388} (1997)
250; D. Tytler et al., astro-ph/9810217 (1998).

\bibitem{quas4} S. Burles and D. Tytler,  Ap.J. {\bf 499} (1998) 699;
Ap.J. {\bf 507} (1998) 732.

\bibitem{cos2} C. Copi, K.A. Olive, and D.N. Schramm, Proc. Nat. Ac.
Sci. {\bf 95} (1998) 2758, astro-ph/9606156.


\bibitem{iz2} Y.I. Izotov, and T.X. Thuan, ApJ, {\bf 500}
(1998) 188.

\bibitem{fdo2} B.D. Fields and K.A. Olive, Ap.J. {\bf 506} (1998)
177.

\bibitem{p} B.E.J. Pagel, E.A. Simonson, R.J. Terlevich and M. Edmunds, 
MNRAS {\bf 255} (1992) 325; E. Skillman, and R.C. Kennicut, ApJ, 411 (1993)
655; E. Skillman, R.J. Terlevich, R.C. Kennicutt, D.R.
Garnett, and E. Terlevich, ApJ, 431(1994) 172.


\bibitem{sfosw}G. Steigman, B. Fields, K.A. Olive, D.N. Schramm, and 
T.P.  Walker, Ap.J. {\bf 415} (1993) L35; M. Lemoine,
D.N. Schramm, J.W. Truran, and C.J. Copi, Ap.J. {\bf 478} (1997) 554;
B.D. Fields and K.A. Olive, astro-ph/9811183, New Astronomy, in press (1998).

\bibitem{rh1} M. Rugers and C.J. Hogan, Ap.J. {\bf 459} (1996)  L1.



\bibitem{kr} L.M. Krauss and P. Romanelli,  ApJ, {\bf 358} (1990) 47.

\bibitem{skm} M. Smith, L. Kawano, and R.A. Malaney, 
Ap.J. Supp., {\bf 85} (1993) 219.

\bibitem{kk1} P.J. Kernan and L.M. Krauss, 
Phys. Rev. Lett. {\bf 72} (1994) 3309.

\bibitem{kk2} L.M. Krauss and P.J. Kernan, Phys. Lett. {\bf B347} (1995) 347.

\bibitem{hata1} N. Hata, R.J. Scherrer, G. Steigman, D. Thomas, and T.P.
Walker, Ap.J., {\bf 458} (1996) 637.

\bibitem{hata2} N. Hata, R. J. Scherrer, G. Steigman, D. Thomas,
T. P. Walker, S. Bludman and P. Langacker, Phys. Rev. Lett. {\bf 75} (1995) 3977.












\bibitem{osb} K.A. Olive and G. Steigman,  Phys. Lett. 
{\bf B354 } (1995) 357.




%\bibitem{spite} F. Spite, and M. Spite,   A.A. {\bf 115} (1982) 357.



\end{thebibliography}
\end{document}